\documentstyle[sprocl]{article}
\def\Journal#1#2#3#4{{#1} {\bf #2}, #3 (#4)}
\def\PRD{{\em Phys. Rev.} D}
\def\vp{\varphi}
\def\al{\alpha}
\def\be{\begin{equation}}
\def\ee{\end{equation}}
\def\D{{\cal D}}
\def\O{{\cal O}}
\def\q{{\bf q}}
\def\gtrsim{\buildrel>\over{\sim}}
\begin{document}
\title{MODELLING THE DECOHERENCE OF SPACETIME}
\author{J.T. WHELAN}
\address{Department of Physics, 201 James Fletcher
Building\\ University of Utah, Salt Lake City, UT 84112}
\maketitle
\abstracts{The question of whether unobserved short-wavelength modes
of the gravitational field can induce decoherence in the
long-wavelength modes (``the decoherence of spacetime'') is addressed
using a scalar field toy model with some features of perturbative
general relativity.  For some long-wavelength coarse grainings, the
Feynman-Vernon influence phase is found to be effective at suppressing
the off-diagonal elements of the decoherence functional.  The
requirement that the short-wavelength modes be in a sufficiently
high-temperature state places limits on the applicability of this
perturbative approach.}

\section{Environment-induced decoherence
in generalized QM}\label{sec:infl}

	To review briefly, the sum-over-histories implementation of
generalized quan\-tum mechanics, as formulated by Hartle\cite{gqm},
assigns to each pair of field histories $\{\vp,\vp'\}$ a {\em
fine-grained decoherence functional}
\be\label{dec}
D[\vp,\vp']=\rho(\vp_i,\vp'_i)\delta(\vp'_f-\vp_f)
e^{i(S[\vp]-S[\vp'])}
\ee
(in units where $\hbar=1=c$), where $\vp_i=\vp(t_i)$ and
$\vp_f=\vp(t_f)$ are the endpoints of the path $\vp$, and similarly
for $\vp'$.  One can {\em coarse grain} by partitioning the field
histories into classes $\{c_\al\}$ and summing the fine-grained
decoherence functional (\ref{dec}) over all field histories lying in a
given class:
\be
D(\al,\al')=\int\limits_\al\D\vp\int\limits_{\al'}\D\vp'\,D[\vp,\vp'].
\ee
If this coarse-grained decoherence functional (which is a Hermitian
matrix with indices $\al$ and $\al'$) is diagonal [$D(\al,{\al
'})\approx\delta_{\al{\al '}}p_\al$], then each diagonal element
$p_\al$ is the probability that the system follows a history in class
$c_\al$.  We say then that the coarse graining $\{c_\al\}$ {\em
decoheres}.  When the alternatives do not decohere, quantum mechanical
interference prevents the theory from assigning  consistent
probabilities (as defined by the classical probability sum rules).

	Most decoherence in physical situations arises\cite{classeq}
when the variables $\vp$ can be split into ``system'' varibles $\Phi$
which can define coarse grainings of interest and and ``environment''
$\phi$ which is only relevant due to its interaction with the system.
The action can then be divided:
$S[\vp]=S_\Phi[\Phi]+S_E[\phi,\Phi]$, into a piece $S_\Phi$ which
would describe the system in the absence of the environment and a
piece $S_E$ which describes the enviroment and its coupling to the
system.  The decoherence functional $D(\al,\al')$ for a coarse
graining based solely on the system variables can be written in terms
of $D[\Phi,\Phi']=\int\D\phi\,\D\phi' D[\vp,\vp']$.
Assuming that the density matrix describing the initial state factors:
\be
\rho(\vp_i,\vp'_i)
=\rho_\Phi(\Phi_i,\Phi'_i)\rho_\phi(\phi_i,\phi'_i),
\ee
then $D[\Phi,\Phi']$ can be written
\be
D[\Phi,\Phi']
=\rho_\Phi(\Phi_i,\Phi'_i)\delta(\Phi'_f-\Phi_f)
e^{i(S_\Phi[\Phi]-S_\Phi[\Phi']+W[\Phi,\Phi'])}
\ee
where
\be
e^{iW[\Phi,\Phi']}=\int\D\phi\,\D\phi' 
\rho_\phi(\phi_i,\phi'_i)\delta(\phi'_f-\phi_f)
e^{i(S_E[\phi,\Phi]-S_E[\phi',\Phi'])}.
\ee
$W[\Phi,\Phi']$ is called the Feynman-Vernon influence
phase \cite{feynvern}; if the \emph{influence functional} $e^{iW}$
becomes small for $\Phi\neq\Phi'$, the ``off-diagonal'' parts of
$D[\Phi,\Phi']$ will be suppressed, causing alternatives defined in
terms of $\Phi$ to decohere.\cite{classeq}

\section{Toy Model}

	The model which we use to mimic general relativity is
that of a scalar field $\vp$ on Minkowski space with action
\be
S[\vp]=-\frac{1}{2}\int d^4\!x\,(\partial_\mu\vp)(\partial^\mu\vp)
[1-(2\pi)^{3/2}\ell\vp].
\ee
This action has the same structure as the first few terms of a
perturbative expansion of GR: a wave term followed by an interaction
term containing two derivatives.  In fact it can be
obtained\cite{dost} from the Nordstr\"{o}m-Einstein-Fokker
theory\cite{NEF} theory of a conformally flat metric
$g_{\mu\nu}=\Omega^2\eta_{\mu\nu}$ by taking $\vp$ proportional to
$\Omega^4-1$ (motivated by the volume element $\sqrt{|g|}=\Omega^4$)
and $\ell$ proportional to the Planck length, and expanding the action
to the first interacting order in $\ell$.

\section{On the applicability of perturbation theory}

As described in section \ref{sec:infl}, decoherence is enforced if
$|e^{iW[\Phi,\Phi']}|\ll1$ whenever $\Phi-\Phi'$ is appreciable.
However, since to zeroth order in perturbation theory, the system and
environment are not coupled, a perturbative expansion will yield
$|e^{iW}|=1+\O(\ell)$.  In order to make $|e^{iW}|$ small, the
$\O(\ell)$ term must be comparable to unity, making this sort
decoherence in some sense an inherently nonperturbative phenomenon.

It is still possible, however, to treat some aspects of the problem
perturbatively if there is another small quantity $\beta$.  In that
case, while terms proportional to $\ell$ can be small and be treated
perturbatively, terms proportional to $\frac{\ell}{\beta}$ can be
large, producing the seemingly nonperturbative effect.  In the present
work, the small parameter is the inverse temperature of the thermal
state assumed to describe the short-wavelength environment.

\section{Results}

If the wavenumber $k_c$ dividing the short- and long-wavelength
regimes and the inverse temperature $\beta$ of the environment obey
$\beta k_c\ll1$, one can calculate\cite{dost} the upper limit
\be
e^{iW[\Phi,\Phi']}\le\frac{1}{(1+A^2[\Delta\Phi])^{1/4}},
\ee
which is indeed small for large values of the suppression factor
\be
A^2\gtrsim\frac{\ell^2}{\beta^2}\int\limits_{q<k_c}d^3\!q
\int_{-q}^q d\omega\,|\Delta\Phi_{\q\omega}|^2
\frac{(2\pi)^2(q^2-\omega^2)^2}{32q|\omega|}
\ln\left(1+\frac{|\omega|}{k_c}\right).
\ee
Considering, for example, a coarse graining by values of a
``field average'' $\langle\Phi\rangle =\Delta\omega(\Delta
q)^3\Phi_{\q_0\omega_0}$, the suppression factor is bounded (when
$q_0\gg\omega_0$) by
\be
A^2\gtrsim\frac{\pi^2}{8}
\frac{|\langle\ell\Delta\Phi\rangle|^2}{\beta^2q_0^2}
\frac{q_0^4}{\Delta\omega(\Delta q)^3}.
\ee
Since the factors $\frac{1}{\beta q_0}$, $\frac{q_0}{\Delta\omega}$,
and $\frac{q_0}{\Delta q}$ are all large, $A^2$ can become large and
suppress the off-diagonal components of $D[\Phi,\Phi']$ even when the
difference $\ell\Delta\Phi=\ell\Phi-\ell\Phi'$ is still small enough
to be within the perturbative regime.

\end{document}